\documentclass{article}
\usepackage{spconf,amsmath,graphicx}

\usepackage{enumitem}
\setlist{nosep, leftmargin=14pt}

\usepackage{mwe} 

\title{Motor Imagery Classification using EEG Spectrograms}
\name{Saadat Ullah Khan$^{1}$, Muhammad Majid$^{1,*}$, Syed Muhammad Anwar$^{2,3}$ \thanks{m.majid@uettaxila.edu.pk}}
\address{$^{1}$ Department of Computer Engineering, University of Engineering and Technology, Taxila, Pakistan.\\$^{2}$ Sheikh Zayed Institute for Pediatric Surgical Innovation, Children’s National Hospital, Washington, DC. \\ $^{3}$ School of Medicine and Health Sciences, George Washington University, Washington, DC.}
\begin{document}
%
\maketitle
\begin{abstract}
The loss of limb motion arising from damage to the spinal cord is a disability that could effect people while performing their day-to-day activities. The restoration of limb movement would enable people with spinal cord injury to interact with their environment more naturally and this is where a brain-computer interface (BCI) system could be beneficial. 
The detection of limb movement imagination (MI) could be significant for such a BCI, where the detected MI can guide the computer system. Using MI detection through electroencephalography (EEG), we can recognize the imagination of movement in a user and translate this into a physical movement. In this paper, we utilize pre-trained deep learning (DL) algorithms for the classification of imagined upper limb movements. We use a publicly available EEG dataset with data representing seven classes of limb movements. We compute the spectrograms of the time series EEG signal and use them as an input to the DL model for MI classification. Our novel approach for the classification of upper limb movements using pre-trained DL algorithms and spectrograms has achieved significantly improved results for seven movement classes. When compared with the recently proposed state-of-the-art methods, our algorithm achieved a significant average accuracy of 84.9\% for classifying seven movements.
\end{abstract}
\section{Introduction}
\label{sec:intro}

Decoding the movement imagination (MI) of users with spinal cord injury (SCI) is gaining relevancy in the current age. Resulting from workplace accidents, falls, road accidents, and injuries during sports people are prone to lose their ability to control their upper limbs due to damage to the spinal cord. Since neural communication between the brain and limbs relies on the spinal cord, any damage to the spinal cord would hinder the flow of biological signals required to perform limb movements \cite{akan2021information}. In such cases, patients could require continuous support from others in the form of financial and/or physical assistance for their daily activities. The use of non-invasive methods such as electroencephalography (EEG) has been getting attention for the restoration of upper limb movement. There are several advantages that such non-invasive methods provide such as simplicity, low cost, safe to use, and convenience when compared to other modalities such as surface electromyography, mechanomyography, electromyography, and functional magnetic resonance imaging \cite{wolpaw2007brain}. Non-invasive EEG is a way to observe and analyze ongoing brain activity and provides time series signals from the scalp using a multi-channel electrode system. When properly utilized, the EEG data could enable people with SCI, through MI detection and robotics, to interact conveniently with their environment.

\begin{figure*}[t]
  \centering
  \centerline{\includegraphics[width = 160 mm]{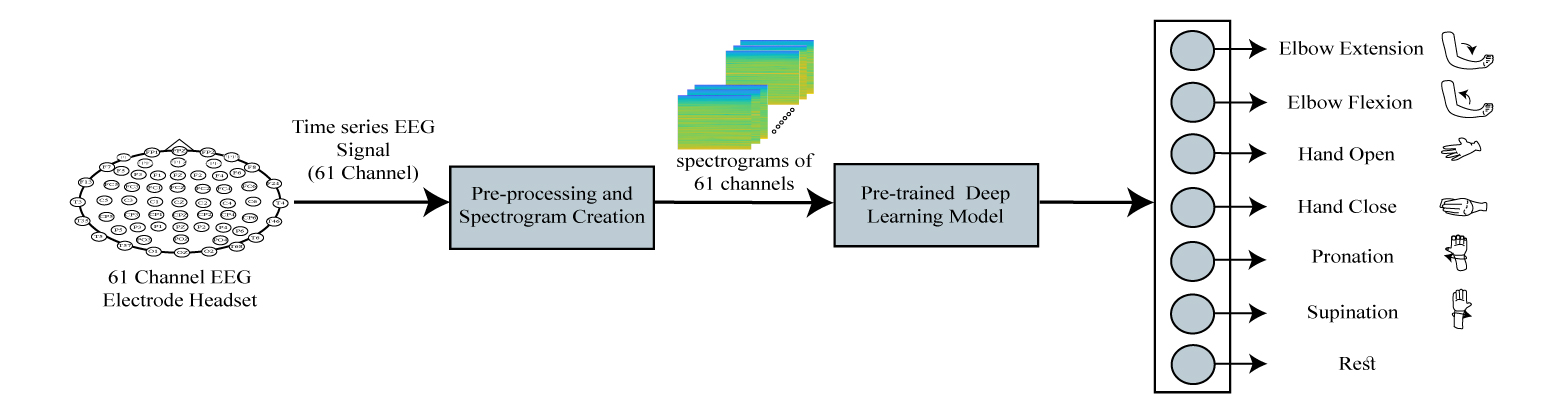}}
  \caption{Our proposed approach for movement imagination classification using EEG spectrograms.}\medskip
  \label{fig:fig1}
\end{figure*}

Recently, several deep learning (DL) models have been proposed that can effectively classify up to four classes of upper limb movements \cite{jiang2020deep}, \cite{jiang2020eeg}, \cite{bressan2021deep}, \cite{valenti2021deep}. A deep convolutional neural network, using a transfer learning paradigm was used to perform binary classification of hand movements \cite{craik2019classification}. A band-power feature refining convolutional neural network 
based on a shallow neural network was developed to classify single-hand movement intention using the band-power features for the classification of four movement classes \cite{lee2020motor}. Both movement execution (ME) and MI tasks were used for the training of a network for the classification of MI tasks using a BCI-transfer learning method based on a relation network 
Features related to ME and MI were extracted from the input EEG data and were combined for four class classification \cite{lee2020decoding}. EEGNet was proposed and used 2-dimensional depth-wise and point-wise convolutional layers for the classification of up to four classes of movement \cite{lawhern2018eegnet}. Deep ConvNet and shallow ConvNet, which comprised of four and two convolution layers respectively, were used for binary classification of hand movements using the raw EEG time series signal as input to the model \cite{schirrmeister2017deep}. 
A variational autoencoder was used to reconstruct the ME signal similar to MI data using the spatial information inherent in the MI signals and has shown some success in four class classification of upper limb movements by using a pre-trained CNN as the classifier \cite{lee2022motor}. In another study, binary classification of MI was performed using a discriminative graph Fourier subspace \cite{miri2022enhanced}. Graph Fourier transform was used on a graph signal constructed from the electrode positioning to yield an optimal representation, which was then used for binary classification of MI tasks.

There is a lack of studies that focus on the inclusion of a diverse set of movements, such as those in the elbow, forearm, and hand. In particular, two or four-movement classification dominates the literature. However, for practical use cases, a diverse set of movements of the elbow, forearm, and hand could help people with SCI translate their MI into a useful physical movement. Further, for the EEG dataset used here, the utility and efficiency of spectrograms have not been evaluated so far to the best of our knowledge. Our proposed study aims to classify seven movement classes (elbow flexion, elbow extension, pronation, supination, hand close, hand open, and rest) that are associated with the elbow, forearm, and hand using the upper limb movement decoding from EEG (001-2017) dataset \cite{ofner2017upper}. 
Towards this, we propose to use spectrograms of the EEG data related to these seven movements. Further, we use two pre-trained DL models to observe the effect on the classification accuracy as the depth of the model is varied. We attain robust classification performance using pre-trained DL algorithms using spectrograms of the EEG signal, achieving the highest classification accuracy (subject wise) of 97.20\% (VGG-16) and 96.74\% (VGG-19) for the deep learning models utilized.

\section{Proposed Methodology}
\label{sec:Method}
Our proposed methodology aims to analyze and classify motion imagination in seven different classes using EEG data. In MI tasks, the user imagines executing the task without any physical movement of their own. To perform the classification of the time series EEG signals, we compute the spectrogram of each channel of the movement and use it as an input to the pre-trained DL models. We use a publicly available dataset, being the only public data that contains seven MI classes. The seven classes that we aim to classify in this study are elbow extension, elbow flexion, hand close, hand open, pronation, supination, and rest. A comprehensive diagram of our proposed methodology is shown in Fig. \ref{fig:fig1}. 

\subsection{EEG Data}
EEG data was recorded using a 61-channel EEG headband at a 512 Hz sampling rate. It contains EEG recordings from $15$ subjects recorded in $10$ trial runs, each run containing $6$ sessions for each movement class.

\subsection{Pre-processing and Spectrogram Generation} 

For the pre-processing of the data, a Chebyshev filter was applied to the raw time series EEG data with the frequency ranging from 0.01 to 200 Hz. Subsequently, a notch filter was applied to reduce the power line interference at 50 Hz. Using this noise-reduced time-series EEG data we compute the spectrogram for each EEG channel.

We use the short-time Fourier transform (STFT) for the calculation of the spectrogram using a sliding window protocol. STFT uses blocks of the original signal to compute the Fourier transform of the signal. The result of this Fourier transform gives us both the time and frequency behavior of the signal \cite{kuisma2005using}. The mathematical representation for STFT is presented in Eq. \ref{eqn:stft}.
\begin{equation}
\label{eqn:stft}
X_{m}(\omega) = \sum_{n=-\infty}^{\infty} x(n + mR) w(n)e^{-\iota\omega(n + mR),}
\end{equation}
where $x$ is the input signal at time $n$, $\omega(n)$ is the window of length $m$, $R$ is the size of hop between the successive DTFTs, and $ X_{m} (\omega)$ is the DTFT of the windowed data.

\subsection{Pre-trained Deep Learning Model}
We use a pre-trained VGG model implemented in PyTorch. It is pre-trained on the ImageNet-1k dataset and is available publicly for use. VGG is a CNN consisting of multiple layers of convolution blocks and fully connected (FC) layers. It uses Relu as the activation function and a small convolutional filter of $3\times 3$ size and a stride of 1-pixel \cite{simonyan2014very}. Here in our study, we use VGG-16 and VGG-19, where VGG-16 consists of 13 convolution layers and VGG-19 consists of 16 convolution layers, with fully connected layers at the end of both these networks. Since the input dimension of the pre-trained VGG is $H \times W \times C$, with $C$ the number of channels required to be three; we duplicate the values of the spectrogram and stacked them on top of one another. For seven class classifications of the data, we replaced the last layer of the pre-trained VGG with a linear layer consisting of seven neurons.

\section{Experimental Results and Discussion}
\label{sec:SR}

We use a publicly available dataset, upper limb movement decoding from EEG (001-2017) \cite{ofner2017upper}. It consists of seven classes and is the only publicly available dataset with seven (highest) movement classes. The movement classes in this dataset are elbow extension, elbow flexion, hand close, hand open, pronation, supination, and rest. It has both ME and MI classes of movement. 


\begin{figure}[t]
  \centering
  \centerline{\includegraphics[width = 90 mm]{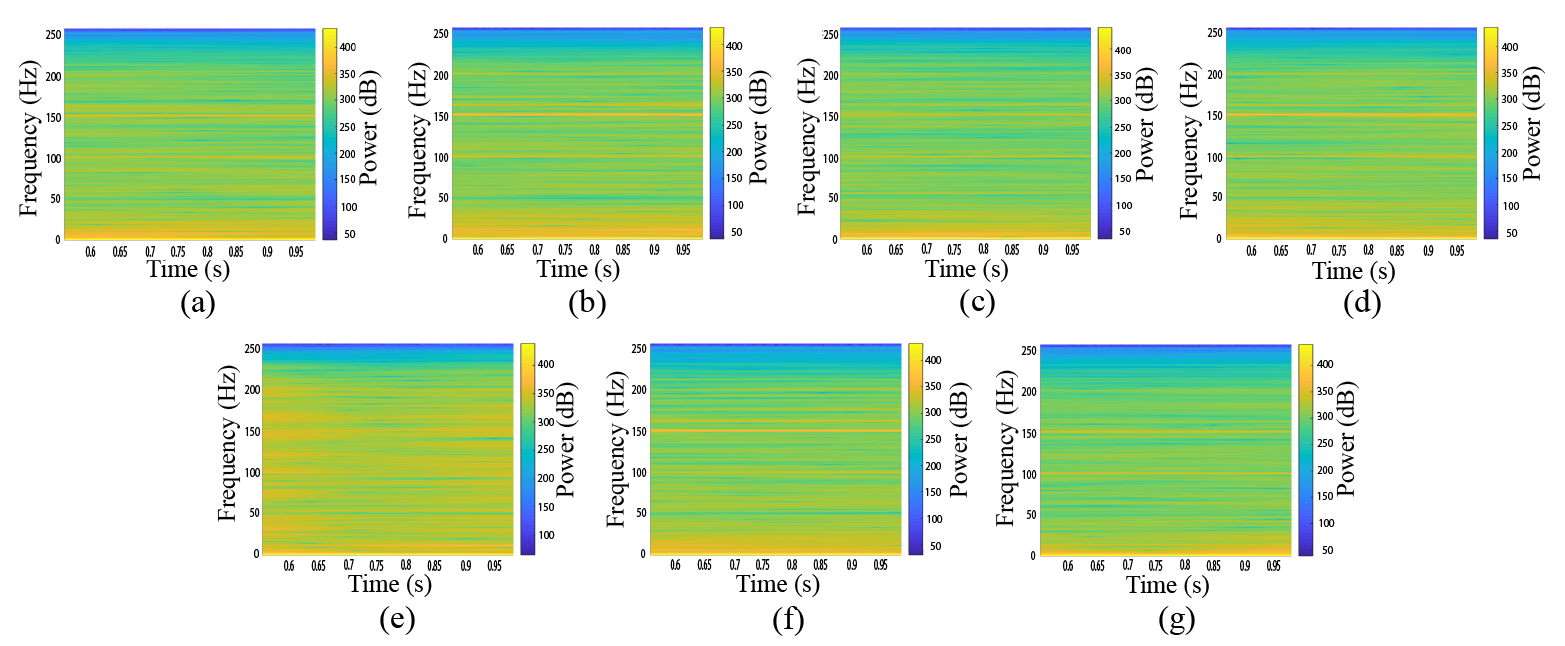}}
  \caption{Spectrograms of different movement imaginations (a) elbow extension, (b) elbow flexion, (c) hand open, (d) hand close, (e) pronation, (f) supination, and (h) rest.}\medskip
  \label{fig:fig2}

\vspace{-5mm}
\end{figure}

As the input size for the pre-trained model is $224\times224$ we calculated the spectrograms of the same size. For this purpose, we took $788$ samples from each trial of the movement and take $447$ points discrete Fourier transform (DFT), which sets the height of the spectrogram at size $224$. 
For the specific width of the spectrogram, we set the window size at $565$ and the number of overlapping samples at $564$. We used the Blackman window throughout the calculation of spectrograms and take the natural log of magnitude square of each of the spectrograms so that, we can get the real values from the computed spectrograms. Though there are many other combinations of the window size and the overlapping window through which spectrogram of size $224\times 224$ can be computed, we have left the study of the classification accuracy of the DL model on different cases of spectrograms to the future work. As our dataset consists of $61$ EEG channels we calculate the spectrogram of each channel independently. For every movement of each subject, there are 10 trial runs and 6 sessions resulting in $3660$ spectrograms for each class for each subject irrespective of the train-valid-test split. Fig. \ref{fig:fig2} presents the spectrogram of channel $FC1$ for the seven classes that are classified in this study. It can easily be observed that the spectrograms are different for each MI class, which helps in better classification of MI from EEG data.  

\begin{table}
\centering
\caption{Classification performance of the proposed method based on EEG spectrogram using VGG-16 and VGG-19 DL models for seven classes of movements.}
\label{tab: T1}
\begin{center}
\begin{tabular}{c c c}
\hline
Subject & VGG-16 & VGG-19 \\
\hline
S1 & 88.93\% & 88.13\%\\
S2 & 82.14\% & 80.34\%\\
S3 & 79.80\% & 76.28\%\\
S4 & 73.06\% & 71.81\%\\
S5 & 76.03\% & 77.79\%\\
S6 & 87.49\% & 87.43\%\\
S7 & 90.80\% & 87.39\%\\
S8 & 96.74\% & 97.20\%\\
S9 & 82.00\% & 83.21\%\\
S10 & 92.29\% & 93.50\%\\
S11 & 91.97\% & 92.07\%\\
S12 & 90.96\% & 88.97\%\\
S13 & 88.03\% & 89.65\%\\
S14 & 91.25\% & 90.41\%\\
S15 & 94.65\% & 94.49\%\\
\hline
Average & 87.07\% & 86.57\% \\
\hline

\vspace{-5mm}
\end{tabular}
\end{center}
\end{table}


The data were divided into $70\%$ for training, $10\%$ for validation, and $20\%$ for testing of the model. The pre-trained DL model (VGG) was fine tuned using $16$ as batch size and for $20$ epochs. The momentum and learning rate was kept at $0.9$ and $0.001$, respectively. We used stochastic gradient descent (SGD) as the optimizer and cross-entropy as the loss function. Specifically, we used VGG with batch normalization otherwise the model would give us an undefined loss after every epoch. The computation of the spectrograms and the training of the DL model was done on a personal computer with Intel\textregistered Xeon\textregistered W-2265 CPU @ 3.50GHz with 64 GB physical memory and Nvidia\textregistered  RTX \textsuperscript{TM} A5000 GPU.

We train the DL model for each subject individually and report the accuracy of the model per subject. Table \ref{tab: T1} shows the performance of VGG-16 and VGG-19 for each of the 15 subjects for all seven classes of movement. For both VGG-16 and VGG-19, we obtain the best classification results for subject 8, achieving an accuracy of $96.74\%$ and $97.20\%$ respectively. Subject 4 performs the worst achieving an accuracy of $73.06\%$ and $71.81\%$ for VGG-16 and VGG-19 respectively.

\begin{table*}
\centering
\caption{Performance comparison of our proposed method with the existing models for 4 MI classification evaluated on the same dataset and subjects. CSP: Common Spatial Pattern, LDA: Linear Discriminant Analysis.}
\label{tab: T2}
\begin{center}
\begin{tabular}{c c c c c c c c}
\hline
Subject & \textbf{Proposed} & \textbf{Proposed} & CVNet  & CSP+LDA  & EEGNet  & Deep  & Shallow \\

 & VGG-16 & VGG-19 &  \cite{lee2022motor} &  \cite{wu2013common} & \cite{lawhern2018eegnet} & ConvNet \cite{schirrmeister2017deep} & ConvNet \cite{schirrmeister2017deep}\\
\hline
S1 & \textbf{87\%} & 84\% & 70\% & 52\% & 62\% & 66\% & 67\%\\
S2 & \textbf{79\%} & \textbf{79\%} & 63\% & 50\% & 56\% & 54\% & 58\%\\
S3 & \textbf{80\%} & 77\% & 72\% & 42\% & 52\% & 62\% & 65\%\\
S4 & \textbf{75\%} & 68\% & 69\% & 44\% & 54\% & 60\% & 52\%\\
S5 & \textbf{80\%} & 75\% & 73\% & 43\% & 52\% & 66\% & 54\%\\
S6 & 86\% & \textbf{89\%} & 72\% & 44\% & 60\% & 67\% & 69\%\\
S7 & 89\% & \textbf{92\%} & 66\% & 47\% & 50\% & 56\% & 54\%\\
S8 & 96\% & \textbf{97\%} & 74\% & 43\% & 58\% & 62\% & 68\%\\
S9 & \textbf{84\%} & 74\% & 63\% & 42\% & 52\% & 54\% & 65\%\\
S10 & \textbf{93\%} & 91\% & 68\% & 46\% & 56\% & 60\% & 62\% \\
\hline
Average & \textbf{84.9}\% & 82.3\% & 69\% & 45.3\% & 55.2\% & 60.7\% & 61.4\%\\
\hline
\end{tabular}
\end{center}
\end{table*}

\begin{figure}[t]
  \centering
  \centerline{\includegraphics[width = 85mm]{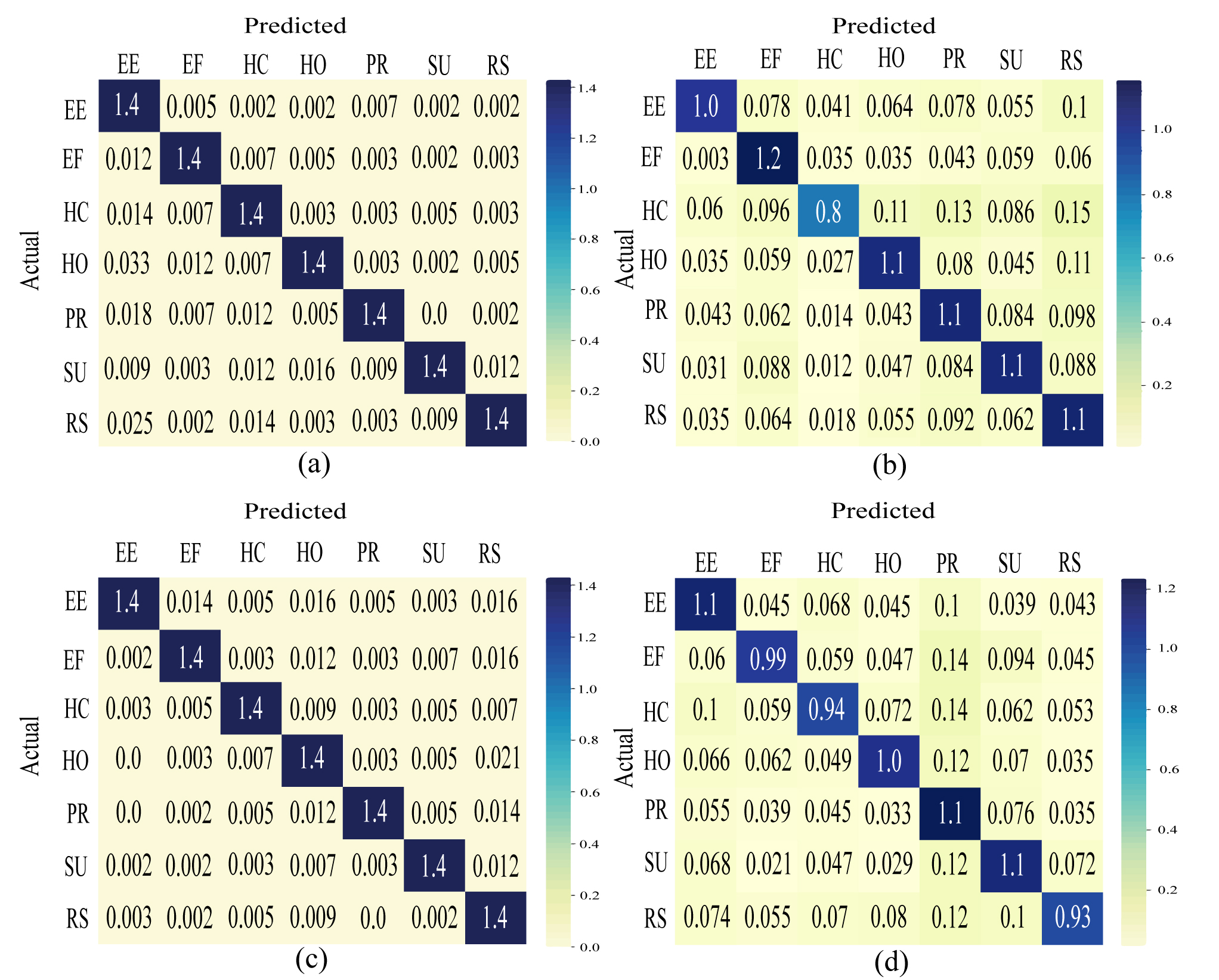}}
  \caption{Confusion matrices for best and worst performing subjects: (a) subject 8 with VGG-16, (b) subject 4 with VGG-16, (c) subject 8 with VGG-19, (d) subject 4 with VGG-19. EE: Elbow Extension, EF: Elbow Flexion, HC: Hand Close, HO: Hand Open, PR: Pronation, SU: Supination, and RS: Rest.}\medskip
  \label{fig:fig4}
  
  \vspace{-7mm}
\end{figure}

Fig. \ref{fig:fig4} represents the confusion matrices for both the best and the worst performing subjects i.e., subjects 8 and 4 respectively for both VGG-16 and VGG-19. It can be seen from the confusion matrix of VGG-16 for subject 8 that it classifies pronation class well, while for subject 4 it shows the worst performance for hand open class. In the case of VGG-19, it shows the best classification performance for hand open, pronation, and rest class. On the other hand, for subject 4, the classification results are not as good as in the former case as VGG-19 shows the least classification performance for the rest and hand close class. We observed that the intraclass variations on the spectrograms for classes showing lower performance in limited and could be the reason for these outcomes. In future we intend to analyze variations in spectrogram generation to further improve the performance. 

For a fair comparison of our proposed method with the models reported in the literature, we apply four class classification of the movement data such that we merge the seven classes into four classes. Particularly, we merged the two elbow classes into a single elbow class, the two hand classes into one hand class, and the two forearm classes into one forearm class. Therefore, we obtain four movement classes for elbow, forearm, hand, and rest. To apply four class classification of our data, we attach a linear layer at the end of the model consisting of four neurons and then train the whole network again for four classes. Table. \ref{tab: T2} shows the classification performance of VGG-16 and VGG-19 with other models which are trained on the same dataset as ours for 4 classes. Our proposed EEG spectrogram-based method outperforms state-of-the-art methods. It should be noted the the comparison results are presented for the 10 subjects used in other studies. For four class movement classifications, we observe that the model with lesser depth (VGG-16) performs slightly better than the deeper model. We argue that since the number of subjects is limited, deeper models are prone to over-fitting and hence a lower test set performance.  
\vspace{-5mm}
\section{Conclusions}
\label{sec:Con}
\vspace{-5mm}
In this work, we proposed the classification of seven upper limb movements based on the spectrogram of the time series EEG signal using pre-trained DL models and a publicly available EEG dataset. 
Such a classification using DL algorithms based on EEG spectrograms for seven classes of movement has not been done before to the best of our knowledge. For seven class movements, we obtained an average classification accuracy of 87.07\% and a maximum accuracy of 97.20\%. Our study has shown that DL models with appropriate spectrogram from each EEG channels show greater classification accuracy than other methods that use feature extraction beforehand, or that use raw EEG as input to models. For four classes, our method improves the accuracy significantly (an improvement of ~15\%).  
In future, the classification accuracy of DL models with different variations of spectrogram generation methods would be evaluated.
\section{Compliance with Ethical Standards}
\vspace{-3mm}
This research study was conducted retrospectively using human subject data made available in open access. Ethical approval was *not* required as confirmed by the license attached with the open access data.
\vspace{-5mm}
\bibliographystyle{IEEEbib}
\bibliography{refs}

\begin{thebibliography}{10}

\bibitem{akan2021information}
Ozgur~Baris Akan, Hamideh Ramezani, Meltem Civas, Oktay Cetinkaya, Bilgesu~Arif
  Bilgin, and Naveed~Ahmed Abbasi,
\newblock ``Information and communication theoretical understanding and
  treatment of spinal cord injuries: State-of-the-art and research
  challenges,''
\newblock {\em IEEE Reviews in Biomedical Engineering}, 2021.

\bibitem{wolpaw2007brain}
Jonathan~R Wolpaw,
\newblock ``Brain-computer interfaces (bcis) for communication and control,''
\newblock in {\em Proceedings of the 9th international ACM SIGACCESS conference
  on Computers and accessibility}, 2007, pp. 1--2.

\bibitem{jiang2020deep}
Yongyu Jiang, Xiaodong Zhang, Chaoyang Chen, Zhufeng Lu, and Yachun Wang,
\newblock ``Deep learning based recognition of hand movement intention eeg in
  patients with spinal cord injury,''
\newblock in {\em 2020 10th Institute of Electrical and Electronics Engineers
  International Conference on Cyber Technology in Automation, Control, and
  Intelligent Systems (CYBER)}. IEEE, 2020, pp. 343--348.

\bibitem{jiang2020eeg}
Yongyu Jiang, Christine Chen, Xiaodong Zhang, Wei Zhou, Chaoyang Chen, and
  Stephen Lemos,
\newblock ``Eeg-based hand motion pattern recognition using deep learning
  network algorithms,''
\newblock in {\em Proceedings of the 2020 9th International Conference on
  Computing and Pattern Recognition}, 2020, pp. 73--79.

\bibitem{bressan2021deep}
Giulia Bressan, Giulia Cisotto, Gernot~R M{\"u}ller-Putz, and Selina~Christin
  Wriessnegger,
\newblock ``Deep learning-based classification of fine hand movements from low
  frequency eeg,''
\newblock {\em Future Internet}, vol. 13, no. 5, pp. 103, 2021.

\bibitem{valenti2021deep}
Andrea Valenti, Michele Barsotti, Davide Bacciu, and Luca Ascari,
\newblock ``A deep classifier for upper-limbs motor anticipation tasks in an
  online bci setting,''
\newblock {\em Bioengineering}, vol. 8, no. 2, pp. 21, 2021.

\bibitem{craik2019classification}
Alexander Craik, Atilla Kilicarslan, and Jose~L Contreras-Vidal,
\newblock ``Classification and transfer learning of eeg during a kinesthetic
  motor imagery task using deep convolutional neural networks,''
\newblock in {\em 2019 41st Annual International Conference of the IEEE
  Engineering in Medicine and Biology Society (EMBC)}. IEEE, 2019, pp.
  3046--3049.

\bibitem{lee2020motor}
Byeong-Hoo Lee, Ji-Hoon Jeong, Kyung-Hwan Shim, and Dong-Joo Kim,
\newblock ``Motor imagery classification of single-arm tasks using
  convolutional neural network based on feature refining,''
\newblock in {\em 2020 8th International Winter Conference on Brain-Computer
  Interface (BCI)}. IEEE, 2020, pp. 1--5.

\bibitem{lee2020decoding}
Do-Yeun Lee, Ji-Hoon Jeong, Kyung-Hwan Shim, and Seong-Whan Lee,
\newblock ``Decoding movement imagination and execution from eeg signals using
  bci-transfer learning method based on relation network,''
\newblock in {\em ICASSP 2020-2020 IEEE International Conference on Acoustics,
  Speech and Signal Processing (ICASSP)}. IEEE, 2020, pp. 1354--1358.

\bibitem{lawhern2018eegnet}
Vernon~J Lawhern, Amelia~J Solon, Nicholas~R Waytowich, Stephen~M Gordon,
  Chou~P Hung, and Brent~J Lance,
\newblock ``Eegnet: a compact convolutional neural network for eeg-based
  brain--computer interfaces,''
\newblock {\em Journal of neural engineering}, vol. 15, no. 5, pp. 056013,
  2018.

\bibitem{schirrmeister2017deep}
Robin~Tibor Schirrmeister, Jost~Tobias Springenberg, Lukas Dominique~Josef
  Fiederer, Martin Glasstetter, Katharina Eggensperger, Michael Tangermann,
  Frank Hutter, Wolfram Burgard, and Tonio Ball,
\newblock ``Deep learning with convolutional neural networks for eeg decoding
  and visualization,''
\newblock {\em Human brain mapping}, vol. 38, no. 11, pp. 5391--5420, 2017.

\bibitem{lee2022motor}
Do-Yeun Lee, Ji-Hoon Jeong, Byeong-Hoo Lee, and Seong-Whan Lee,
\newblock ``Motor imagery classification using inter-task transfer learning via
  a channel-wise variational autoencoder-based convolutional neural network,''
\newblock {\em IEEE Transactions on Neural Systems and Rehabilitation
  Engineering}, vol. 30, pp. 226--237, 2022.

\bibitem{miri2022enhanced}
Maliheh Miri, Vahid Abootalebi, and Hamid Behjat,
\newblock ``Enhanced motor imagery-based eeg classification using a
  discriminative graph fourier subspace,''
\newblock in {\em 2022 IEEE 19th International Symposium on Biomedical Imaging
  (ISBI)}. IEEE, 2022, pp. 1--5.

\bibitem{ofner2017upper}
Patrick Ofner, Andreas Schwarz, Joana Pereira, and Gernot~R M{\"u}ller-Putz,
\newblock ``Upper limb movements can be decoded from the time-domain of
  low-frequency eeg,''
\newblock {\em PloS one}, vol. 12, no. 8, pp. e0182578, 2017.

\bibitem{kuisma2005using}
M~Kuisma and P~Silventoinen,
\newblock ``Using spectrograms in emi-analysis-an overview,''
\newblock in {\em Twentieth Annual IEEE Applied Power Electronics Conference
  and Exposition, 2005. APEC 2005.} IEEE, 2005, vol.~3, pp. 1953--1958.

\bibitem{simonyan2014very}
Karen Simonyan and Andrew Zisserman,
\newblock ``Very deep convolutional networks for large-scale image
  recognition,''
\newblock {\em arXiv preprint arXiv:1409.1556}, 2014.

\bibitem{wu2013common}
Shang-Lin Wu, Chun-Wei Wu, Nikhil~R Pal, Chih-Yu Chen, Shi-An Chen, and
  Chin-Teng Lin,
\newblock ``Common spatial pattern and linear discriminant analysis for motor
  imagery classification,''
\newblock in {\em 2013 IEEE Symposium on Computational Intelligence, Cognitive
  Algorithms, Mind, and Brain (CCMB)}. IEEE, 2013, pp. 146--151.

\end{thebibliography}

\end{document}